\documentclass[aps,prb,reprint,superscriptaddress]{revtex4-2}

\usepackage[utf8]{inputenc}

\usepackage[hidelinks]{hyperref}

\usepackage{hyperref}
\usepackage{color}
\hypersetup{colorlinks=true}

\makeatletter
\@addtoreset{equation}{myequation}
\@addtoreset{figure}{myfigure}
\@addtoreset{table}{mytable}
\makeatother
\usepackage{graphicx}
\usepackage{bm}
\usepackage{amsmath}
\usepackage{amssymb}

\usepackage{physics}
\usepackage{siunitx}
\usepackage{chemformula}

\newcommand{\kv}{\bm{k}}

\newcommand{\rv}{\bm{r}}

\newcommand{\alphaIF}{\alpha_{\text{IF}}}
\DeclareMathOperator{\diag}{diag}
\newcommand{\Hangle}{\theta_{H}}
\newcommand{\gv}{\bm{g}}
\graphicspath{{./figures/}}

\newcommand{\mm}{\mathrm{mm}}
\newcommand{\mr}{\mathrm{mr}}

\newcommand{\bcurv}{\Omega}

\DeclareMathOperator{\erfc}{Erfc}

\begin{document}


\title{Theory for electrical detection of the magnon Hall effect induced by dipolar interactions}

\author{Pieter M. Gunnink}
\email{p.m.gunnink@uu.nl}
\affiliation{Institute for Theoretical Physics and Center for Extreme Matter and Emergent Phenomena, Utrecht University, Leuvenlaan 4, 3584 CE Utrecht, The Netherlands}

\author{Rembert A. Duine}
\affiliation{Institute for Theoretical Physics and Center for Extreme Matter and Emergent Phenomena, Utrecht University, Leuvenlaan 4, 3584 CE Utrecht, The Netherlands}
\affiliation{Department of Applied Physics, Eindhoven University of Technology, P.O. Box 513, 5600 MB Eindhoven, The Netherlands}
\author{Andreas R\"uckriegel}
\affiliation{Institut f\"ur Theoretische Physik, Universit\"at Frankfurt, Max-von-Laue Strasse 1, 60438 Frankfurt, Germany}

\date{\today}

\begin{abstract}
We derive the anomalous Hall contributions arising from dipolar interactions to diffusive spin transport in magnetic insulators. Magnons, the carriers of angular momentum in these systems, are shown to have a non-zero Berry curvature, resulting in a measurable Hall effect. For yttrium iron garnet (YIG) thin films we calculate both the anomalous and magnon spin conductivities. We show that for a magnetic field perpendicular to the film the anomalous Hall conductivity is finite. This results in a non-zero Hall signal, which can be measured experimentally using Permalloy strips arranged like a Hall bar on top of the YIG thin film. We show that electrical detection and injection of spin is possible, by solving the resulting diffusion-relaxation equation for a Hall bar. We predict the experimentally  measurable Hall coefficient for a range of temperatures and magnetic field strengths. Most strikingly, we show that there is a sign change of the Hall coefficient associated with increasing the thickness of the film.
\end{abstract}

\maketitle

\section{Introduction}

One of the earliest successes of the concepts of geometry and topology in condensed matter was the explanation of the anomalous Hall effect in terms of the Berry phase. The anomalous Hall effect was therefore a stepping stone for further understanding of geometrical and topological effects, such as the quantum Hall effect \cite{nagaosaAnomalousHallEffect2010}. Since it is a geometrical effect, the anomalous Hall effect is not restricted to electronic systems. Indeed, it has also been observed for other types of carriers, such as phonons and photons \cite{qinBerryCurvaturePhonon2012, strohmPhenomenologicalEvidencePhonon2005,onodaHallEffectLight2004}. Since spin waves, or magnons, are the carriers of angular momentum in ferromagnets, the question thus naturally arises {if a magnon analogue of the anomalous Hall effect can also exist.} Continuing the analogy with the anomalous Hall effect, the magnon Hall effect could lead to further understanding of topology in magnonic systems.

Previously, a thermal magnon Hall effect has been proposed, where magnons are the heat carriers. First predicted for chiral quantum magnets \cite{katsuraTheoryThermalHall2010}, it was subsequently observed in \ch{Lu2V2O7} \cite{onoseObservationMagnonHall2010,ideueEffectLatticeGeometry2012}. In these systems the chiral nature of the spin waves provides the time-reversal symmetry breaking that is necessary for a finite anomalous Hall response.
For forward volume magnetostatic spin waves in a thin-film ferromagnet a thermal magnon Hall effect has also been proposed \cite{matsumotoTheoreticalPredictionRotating2011,matsumotoRotationalMotionMagnons2011}, where the dipole-dipole interaction provides the required symmetry breaking. A transverse thermal Hall conductivity has also been calculated for this system \cite{matsumotoThermalHallEffect2014}, but has not yet been measured experimentally. This is most likely due to the small transverse thermal conductivities predicted for the most commonly used insulating ferromagnet, yittrium iron garnet (YIG) \cite{onoseObservationMagnonHall2010}. Moreover, phonons also contribute to the thermal Hall effect, and it might therefore be hard to disentangle the contributions of the two heat carriers. An effort has been made by Tanabe \textit{et al.} \cite{tanabeObservationMagnonHalllike2016} to excite spin waves using a coplanar waveguide and measure the temperature gradient perpendicular to the propagation direction. However, they were only able to measure a transverse temperature gradient in the unsaturated regime, which can therefore not directly be attributed to magnons.

\begin{figure}
	\centering
	\includegraphics[width=\columnwidth]{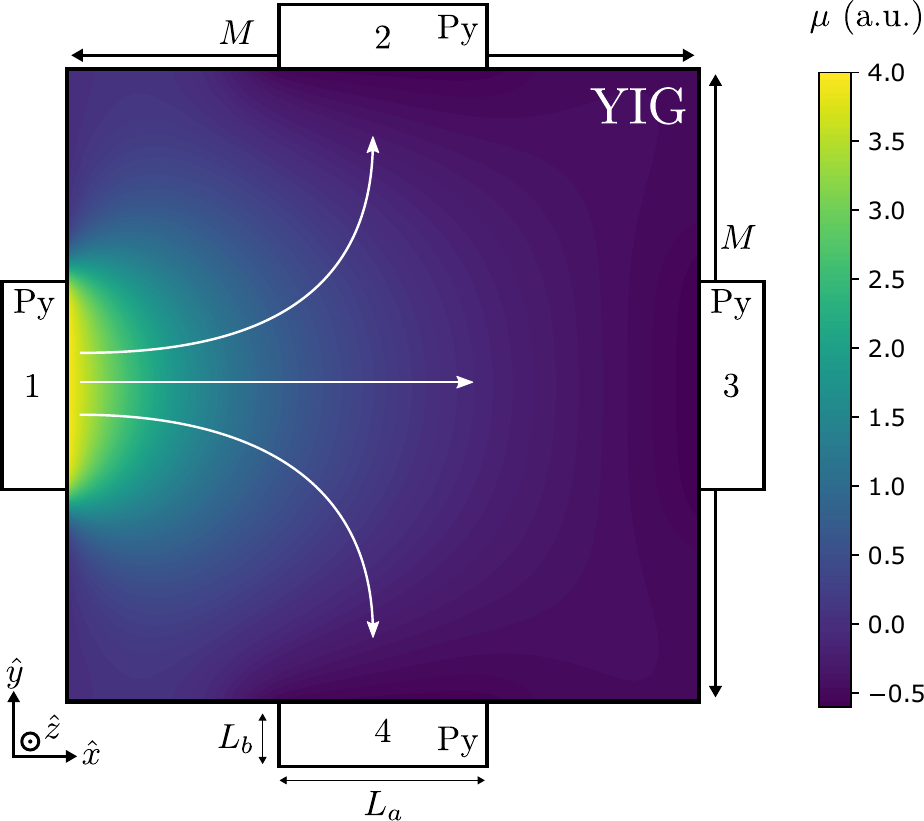}
	\caption{The Hall bar with electrical injection and detection of spin currents using Permalloy strips on top of YIG. Spin current is injected by the Py strip 1, and is detected by the strips 2, 3 and 4. The colorscale shows the diffusion of the magnon chemical potential throughout the film, obtained by solving the diffusion-relaxation equations as described in Sec.~\ref{sec:diffusion}.
          The Hall bar has size $M\times M$, and the Py detectors and detectors have size $L_a\times L_b$, where $L_b\ll L_a$. The magnetic field is oriented out of plane, as shown in Fig.~\ref{fig:geometry}, where also the interface between the YIG and the Py is shown in more detail. 
		\label{fig:hall-bar}}
\end{figure}

{Recent advances have shown that it is possible to electrically inject and detect spin waves using metallic leads} \cite{cornelissenLongdistanceTransportMagnon2015}. This has opened the way to electrically measure the magnon Hall effect. However, a complete picture of the interaction between the electrical detection and the Hall effect is still lacking. Electrical detection via metal strips can significantly modify magnon transport properties \cite{cornelissenSpinCurrentControlledModulationMagnon2018}, and it is not clear if a finite magnon Hall response can still survive. {In this work we therefore develop a theory for the electrical detection of the magnon Hall effect in order to determine if the magnon Hall effect can be measured electrically.}

We numerically calculate the Hall response, using the diffusion-relaxation equation for magnons in a Hall bar geometry, as depicted in Fig.~\ref{fig:hall-bar}. In order to determine the magnitude of the expected Hall response two contributing factors need to be calculated: (1) the magnon spin and anomalous conductivities and (2) the boundary conditions which incorporate the electrical detection. We numerically calculate these using a microscopic description. Starting from the Keldysh quantum kinetic equations \cite{kamenevKeldyshTechniqueNonlinear2009}, we derive the equation of motion of the magnon distribution function to leading order in a semiclassical expansion in gradients. This allows us to separate the spin diffusion and anomalous Hall contributions to the spin current.

This work is ordered as follows. We first discuss the specific Hall geometry required to measure a finite magnon Hall effect in Sec.~\ref{sec:setup}. Next, in order to determine the magnitude of the magnon Hall effect we derive the equations of motion for the spin density in Sec.~\ref{sec:qke}. We also show how the equations of motion have to be modified if a metallic lead is interfaced with the system, in order to detect or inject spins. From the equation of motion we derive a diffusion-relaxation equation, which fully describes the magnon diffusion and relaxation in the Hall bar geometry, including boundary conditions. In Sec.~\ref{sec:coefficients} we show how the conductivities and damping can be numerically evaluated and we discuss results for a typical thin film of YIG. In Sec.~\ref{sec:results} we solve the diffusion-relaxation equation numerically and present our results for a YIG Hall bar, where spin waves are injected and detected electrically. A summary and conclusion are given in Sec.~\ref{sec:conclusions}. In the appendices \ref{app:qke}-\ref{app:hamiltonian} we give a more detailed derivation of the quantum kinetic equations for general bosonic systems, and more details regarding the diffusion-relaxation equation and the Hamiltonian.

\section{Setup}
\label{sec:setup}
First, we discuss the experimental setup necessary to measure a magnon Hall effect electrically. We consider a Hall bar geometry, as shown in Fig.~\ref{fig:hall-bar}. There are four terminals, formed by metal strips on top of a YIG thin film. The strips act as injectors and detectors of spin currents. Magnons are injected at terminal 1 and diffuse through the film. They are then detected at terminals 2, 3 and 4. By comparing the detected currents at terminals 2 and 4 a Hall signal can be measured. Note that in electronic Hall experiments terminal 3 is necessary in order for a current to flow, but in our case we have only included it for completeness.

\begin{figure}
	\centering
	\includegraphics[width=\columnwidth]{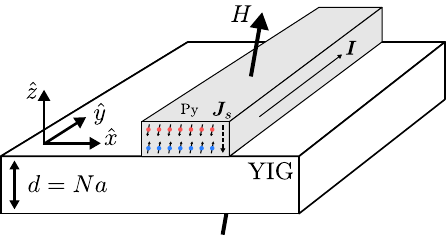}
	\caption{The considered geometry, with the magnetic field pointing slightly off the $\hat{z}$-axis, as explained in the main text. The Py strip on top of the YIG has a charge current $\bm{I}$ running parallel to the film, which induces a spin current $\bm{J}_s$ such that there is an accumulation of spin at the interface between the YIG and the Py. 
		\label{fig:geometry}}
\end{figure}

The Berry curvature is only non-zero if either time-reversal or inversion symmetry is broken \cite{xiaoBerryPhaseEffects2010}. Breaking these symmetries can be achieved by applying a magnetic field perpendicular to the plane, which leads to forward volume modes, as was previously suggested by Matsumoto and Murakami \cite{matsumotoTheoreticalPredictionRotating2011}. Conventionally, one would use the spin Hall effect (SHE) in the metal strips to excite magnons in the YIG film \cite{cornelissenLongdistanceTransportMagnon2015}. {However, the polarization of the spin current induced by the SHE is always in-plane} \cite{hirschSpinHallEffect1999} {and can therefore not excite forward-volume modes in the YIG film. }Instead, we propose to use ferromagnetic Permalloy (Py) strips. If a charge current flows through the Py strip, the anomalous spin Hall effect (ASHE) induces a spin current polarized parallel to the magnetization of the Py strip, as shown in Fig.~\ref{fig:geometry}. For sufficiently large external magnetic fields the magnetization of the Py strips and the YIG will both be aligned to the external field.
{This spin current can therefore excite magnons in the YIG film.} However, the spatial direction of the spin current is $\bm{J}_s\sim\bm{I}\times \bm{M}$, where $\bm{I}$ and $\bm{M}$ are the charge current and magnetization respectively \cite{dasSpinInjectionDetection2017,dasEfficientInjectionDetection2018}. Therefore, if the magnetic field is oriented along the $\hat{z}$ direction and the charge current flows along the $\hat{y}$ direction the spin current flows along the $\hat{x}$ direction. In other words, {the spin current in the Py strip flows parallel to the YIG film and can therefore not enter it to excite magnons}. However, one can tilt the magnetic field slightly off-axis, i.e. off the $\hat{z}$-axis, as depicted in Fig.~\ref{fig:geometry}. The spin current induced by the ASHE then gains an out-of-plane component and {is able to excite magnons in the YIG} \cite{dasEfficientInjectionDetection2018}. At the detectors the opposite process, the inverse ASHE, converts a spin current in a measurable charge current.

\section{Method}
\label{sec:qke}
In this section we consider the microscopic Hamiltonian for a thin film of YIG and derive the equations of motion for the spin density. The formalism that we use, however, is completely general and can be applied to any bosonic Hamiltonian with anomalous coefficients.

We consider a thin film of YIG, with $N$ layers, of thickness $d=Na$, with a magnetic field perpendicular to the film. 
{We include both the dipole-dipole and exchange interaction, which gives us a full description of the spin wave dynamics.} We apply a Holstein-Primakoff transformation to the Hamiltonian, retain terms up to second order, and Fourier-transform along the $x,y$ directions. We can then write the quadratic part of the Hamiltonian as
\begin{equation}
{\mathcal{H}}_{\kv} = \sum_{\kv}
\begin{pmatrix}
 	\bm{b}_{\kv}^\dagger & \bm{b}_{\kv}
\end{pmatrix}
\begin{pmatrix}
\bm{A}_{\kv} & \bm{B}_{\kv}\\
\bm{B}_{\kv}^{\dagger} & \bm{A}_{\kv}
\end{pmatrix}
\begin{pmatrix}
\bm{b}_{\kv}\\
\bm{b}^\dagger_{-\kv}
\end{pmatrix},
  \end{equation}      
where $\bm{b}^\dagger_{\kv} = (  b_{\kv}^\dagger( z_1),...,b^\dagger_{\kv}( z_N))$ are the creation operators for magnons with the two-dimensional wave vector $\kv$ and $\bm{A}_{\kv}$ and $\bm{B}_{\kv}$ are $N\times N$ matrices with $N$ the number of internal degrees of freedom within a unit cell, which is in our case equivalent to the number of layers. More details are found in Appendix~\ref{app:hamiltonian}.
We evaluate the dipole-dipole interaction using the Ewald summation method \cite{kreiselMicroscopicSpinwaveTheory2009}. This allows us to accurately compute the magnon spectrum, even at long wavelengths, where conventional summing methods are slow \cite{ costafilhoMicroscopicTheoryDipoleexchange2000}, but where we do expect the Berry curvature to be large \cite{matsumotoRotationalMotionMagnons2011}. 
From the anomalous coefficients $\bm{B}_{\kv}$, which are due to the dipole-dipole interaction, it is clear that spin is not conserved. The dipolar interactions couple the magnons the the lattice, which therefore acts as a spin sink and/or source. 

We note that the anomalous coefficients in the Hamiltonian create a squeezed magnon state, which is not an eigenstate of the spin in the $z$-direction \cite{kamraMagnonsqueezingNicheQuantum2020}. Between a metallic lead and the magnetic system there is thus an interface of a squeezed (the YIG) and a spin state with definite spin in the $z$-direction (the metallic lead). {This leads to corrections to the spin current over the interface}, which we show in more detail in Sec.~\ref{sec:metalliclead}.

In a bosonic system with anomalous coefficients, the Bogoliubov-de Gennes (BdG) Hamiltonian $\mathcal{H}_{\kv}$ is diagonalized by a para-unitary transformation \cite{colpaDiagonalizationQuadraticBoson1978}, such that 
\begin{equation}
\mathcal{T}^\dagger_{\kv}\mathcal{H}_{\kv} \mathcal{T}_{\kv} = 
	\mathcal{E}_{\kv}
;\quad \mathcal{T}_{\kv}^\dagger\nu \mathcal{T}_{\kv} = \nu,
\end{equation}
where $\mathcal{E}_{\kv}=\diag\left[E_{\kv}^1,...,E_{\kv}^N,E_{-\kv}^1,...,E_{-\kv}^N\right]$ , $\nu=\diag\left[1,...,1,-1,...,-1\right]$ and $\mathcal{T}_{\kv}$ is a para-unitary transformation matrix of size $2N\times 2N$. Note that we only have $N$ distinct bands, since the bands $n$ and $n+N$ are related to each other via the para-unitary structure.

{In order to derive the equations of motion we perform the gradient expansion of the Hamiltonian.} We first define the Berry connection (suppressing the $\kv$-label from here onwards) 
\begin{equation}
	A^{\alpha}=i\nu \mathcal{T}^{\dagger}\nu\left(\partial_{k_{\alpha}}\mathcal{T}\right),
\end{equation}
where $\alpha\in(x,y)$. Numerically, we calculate the Berry connection using the component-wise form
\begin{equation}
	\mathcal{A}_{nm}^{\alpha}=-i\frac{\left[\mathcal{T}^{\dagger}\left(\partial_{k_{\alpha}}H\right)\mathcal{T}\right]_{nm}}{\mathcal{E}_{n}-\nu_{n}\nu_{m}\mathcal{E}_{m}},\quad n\neq m,
	\label{eq:berryphase}
\end{equation}
where $n,m=1,...,2N$. This form also makes it clear that the Berry connection increases close to band crossings. 

From the Berry connection we define the Berry curvature for the $n$-th band as
\begin{align}
	\bcurv_{n}^{\alpha\beta} &= \left( \partial_{k_{\alpha}} A^\beta - \partial_{k_\beta} A^\alpha \right)_{nn} \nonumber \\
	&= i\left( A^\alpha A^\beta - A^\beta A^\alpha \right)_{nn} \label{eq:berrycurvature}
\end{align}
The Berry curvature satisfies the sum rule $\sum_n \bcurv_{n}^{\alpha\beta}=0$, where $n$ is summed over all $2N$ bands.
We note that these definitions for the Berry phase and curvature are equivalent to those given by Shindou \textit{et al.} \cite{shindouTopologicalChiralMagnonic2013}, who were the first to consider the topology of magnons, and also to those of Lein and Sato \cite{leinKreinSchrodingerFormalismBosonic2019}, who showed rigorously that the concept of the Berry phase can be applied to BdG-type Hamiltonians.

{Now we are able to derive the equations of motions for general bosonic systems with non-zero anomalous coefficients.} As noted, this is applicable to the magnons described here, but also for other bosonic systems, such as phonons and photons \cite{onodaHallEffectLight2004,luTopologicalPhotonics2014}, where geometrical effects are also known.
We start from the quantum kinetic equations in the Keldysh formalism, which are derived by performing a Wigner transformation and expanding the gradients up to first order \cite{kamenevKeldyshTechniqueNonlinear2009}. Moreover, we assume damped quasiparticles in (local) thermal equilibrium. We have relegated the details of this calculation to Appendix \ref{app:qke} and will only state the equation of motion for the spin density $s^z(\rv,t)$  here, which is given by

\begin{equation}
	\partial_t s^z + \nabla \cdot \bm{J}_s = \Gamma_s \mu_m,
	\label{eq:eq_motion}
\end{equation}
where we have only kept terms up to first order in the magnon chemical potential $\mu_m$. Here, $\Gamma_s$ describes the relaxation rate of the magnons. The spin current $\bm{J}_s$ is written component-wise as
\begin{equation}
	J_s^\alpha= \sigma_s\partial_{r_{\alpha}}\mu_m+\sigma_s^H\sum_\beta\varepsilon^{\alpha\beta}\partial_{r_{\beta}}\mu_m,
	\label{eq:current}
      \end{equation}      
where $\sigma_s$ is the magnon spin conductivity, $\sigma_s^H$ is the Hall conductivity and $\varepsilon^{\alpha\beta}$ is the two-dimensional Levi-Civita symbol. The Berry curvature only affects the magnon Hall conductivity $\sigma_s^H$, and bands with a greater Berry curvature contribute to a larger Hall conductivity. From the Keldysh formalism the coefficients $\sigma_s, \sigma_s^H$ and $\Gamma_s$ can be calculated using the microscopic Hamiltonian, by integrating the relevant quantities over the entire Brillouin zone. We show the details of this calculation in Appendix \ref{app:hamiltonian}. We consider a clean system in the low-temperature limit, such that the dominant damping source is the Gilbert damping \cite{benderDynamicPhaseDiagram2014}. Moreover, we disregard heat transport, since long-range magnon transport is dominated by the magnon chemical potential \cite{cornelissenMagnonSpinTransport2016}.

The complete magnon dynamics are thus given by Eq.~(\ref{eq:eq_motion}), where we calculate the transport coefficients using the microscopic Hamiltonian. {We therefore do not have to rely on fitting parameters.}

\subsection{Metallic lead}
\label{sec:metalliclead}
{In order to model the electrical detection and injection, we consider a metallic lead interfaced with the YIG film,} as shown in Fig.~\ref{fig:geometry}.
As a result of this interface the equations of motion have to be modified, such that we have at the interface between the magnet and the metallic lead that
\begin{equation}
	\partial_t s^z \left(\boldsymbol{r},t\right) + \nabla \cdot \bm{J}_s  = \Gamma_s \mu_m + A\mu_m + B\mu_e + C,
	\label{eq:spin-sink-diffusion}
\end{equation}
where $\mu_e$ is the electron spin accumulation in the lead. We show the detailed derivation of this correction and the coefficients $A, B$ and $C$ in Appendix~\ref{app:spinsink}. The correction $A\mu_m$, with $A>0$, describes the relaxation of the magnons into the metallic lead. $B\mu_e$ is the injection of spin driven by the chemical potential in the metallic lead. 

The constant $C$ is related to the fact that the magnons are squeezed, whereas the spins in the metallic lead are not squeezed. The main correction is a constant injection of angular momentum into the YIG, even with zero chemical potential in the Py lead, which is a characteristic feature of elliptic magnonic systems \cite{ruckriegelHannayAnglesMagnetic2020}.
The source of this spin current is the lattice, which couples to the magnons via the dipole-dipole interaction. The constant $C$ is therefore zero in the absence of dipolar interactions. There are also corrections due to dipolar interactions to the constants $A$ and $B$, which are of less importance. In absence of these corrections we would have $A=-B$, such that the spin current is zero when $\mu_e=\mu_m$ \cite{cornelissenMagnonSpinTransport2016}.
{With the metallic lead modelled, we now have all the necessary parts for a full description of the dynamics of magnons in a Hall bar.}
\subsection{Diffusion-relaxation}
\label{sec:diffusion}
We now write down the full diffusion-relaxation equation, which we solve numerically to give the full description of the Hall bar, including electrical injection and detection. 
Since the Hall conductivities enter through antisymmetric terms in the current, see Eq.~(\ref{eq:current}), these drop out in the final diffusion-relaxation equation, which becomes 
\begin{equation}
	\sigma_s \nabla^2\mu_m=\Gamma_s\mu_m.
	\label{eq:diffusion}
\end{equation} 
The Hall conductivities only appear in the expressions for the boundary conditions, where we require that the normal component of the current vanishes, i.e. that $\bm{J}_s\cdot\hat{\bm{n}}  = 0$ at the edges of the film if there is no metallic lead present, where $\hat{\bm{n}}$ is the normal vector to the boundary. To measure a finite Hall response we consider a Hall bar setup, as shown in Fig.~\ref{fig:hall-bar}. The Hall response can then be measured between terminals 2 and 4. As far as we are aware, there are no analytical solutions for such a geometry. {We therefore numerically solve the diffusion-relaxation equation, Eq.}~(\ref{eq:diffusion}).

Specifically, we solve the diffusion-relaxation equation on the square $0 \leq x \leq M$ and $0 \leq y \leq M$, where the diffusion is given by Eq.~(\ref{eq:diffusion}). We use a Finite Element Method, with a symmetric square grid, implemented in the FreeFEM++ software \cite{hechtNewDevelopmentFreefem2012}. At the open boundaries we require that $\bm{J}_s \cdot \hat{\bm{n}} = 0$. At the injector and detectors we have the boundary condition $\bm{J}_s \cdot \hat{\bm{n}} = J^{\text{int}}_s(\mu_m)$, where the interface current $J^{\text{int}}_s$ is a function of the magnon chemical potential at the interface $\mu_m^{int}$ and includes the contributions $A,B$ and $C$ as discussed in Sec.~\ref{sec:metalliclead}. We give the full form of $J^{\text{int}}_s$ in Appendix~\ref{app:bc}. 
We then define the total spin current injected or detected at Py strip $i$ as 
$I_{i}=\int_{\partial S_i} \bm{J}_s\cdot \hat{\bm{n}} \dd{s} $, 
where $\partial S_i$ is the interface between the Py and the YIG.

\section{Hall angle and diffusion length}
\label{sec:coefficients}

With the full description of the transport coefficients complete, we now numerically evaluate these using the microscopic Hamiltonian. We have relegated the derivation of these coefficients to Appendix~\ref{app:coefficents}.  The parameters used in this work are shown in Table~\ref{tab:values}. We only consider the low-temperature regime $T<\SI{2}{K}$, since at higher temperatures we expect other damping mechanisms besides the Gilbert damping to play a role. Moreover, one might expect the ferrimagnetic branches in the YIG dispersion relation to be relevant at room temperature \cite{barkerThermalSpinDynamics2016}, which are not captured in our model.

\begin{figure}
	\centering
	\includegraphics{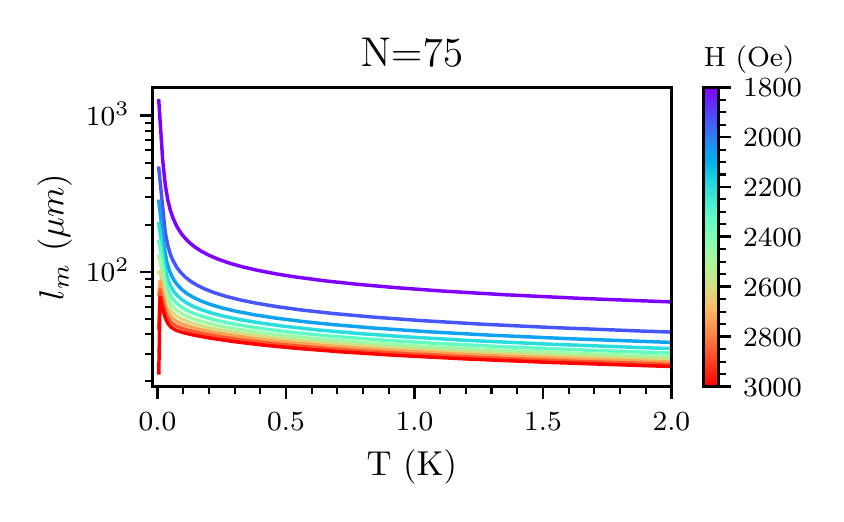}
	\caption{The spin diffusion length $l_m$ for a thin film of YIG with thickness N=75, for varying magnetic field strength. The corresponding Hall angle is shown in Fig.~\hyperref[fig:compare75_150]{\ref*{fig:compare75_150}a}.  
		\label{fig:diffusionlength}}
\end{figure}

{First, we show the results for the spin diffusion length,} $\ell_m=\sqrt{\sigma_s/\Gamma_s}$, for a film of thickness $N=75$ in Fig.~\ref{fig:diffusionlength}. The diffusion length peaks for low temperatures, and converges to a constant value in the high temperature regime. This can be explained by the energy dependence of the Gilbert damping: for low temperature only the lowest energy bands contribute, which have the lowest Gilbert damping, since the damping is proportional to energy. The drop-off of the diffusion length at low temperature and high magnetic field is explained by the fact that the temperature is not high enough to occupy the first band, and there is thus no transport possible. At elevated temperatures we compare the spin diffusion length to a simple model that only considers the lowest exchange band of YIG, from which the spin diffusion length is estimated as $l_m\approx 4\sqrt{J / 3k_B T M_s \alpha_G^2}$ \cite{cornelissenMagnonSpinTransport2016}.  We expect this approximation to be only valid for relatively high temperatures, where the higher exchange bands are occupied, and for thicker films. We therefore compare this approximation with our calculations at $T=\SI{2}{K}$ and find that $l_m\approx \SI{35}{\micro m}$, whereas our numerical model found $l_m=\SI{55}{\micro m}$ for $N=150$ and $H=\SI{1800}{Oe}$. Moreover, as is evident from Fig.~\ref{fig:diffusionlength}, our numerically calculated diffusion length also scales as $1/\sqrt{T}$. For different thicknesses (not shown here) the behaviour and order of magnitude of the spin diffusion length is similar.

\begin{table}[]
\centering
\caption{Parameters for YIG used in the numerical calculations in this work. Note that $S$ follows from $S=M_sa^3/\mu$, where $\mu=2\mu_B$ is the magnetic moment of the spins, with $\mu_B$ the Bohr magneton. We are not aware of any values of the parameters $\mu_e$ and $\alpha_{IF}$ for a YIG\textbar Py interface and have therefore assumed values that are equivalent to the YIG\textbar Platinum interface. Since the injection and detection is described in linear response, their exact values do not affect the final results.  \label{tab:values}}
\begin{ruledtabular}
	\begin{tabular}{@{\hspace{6em}} lc @{\hspace{6em}}}
		Quantity & Value \rule{0pt}{2.6ex}\rule[-1.2ex]{0pt}{0pt}\\
		$a$ &  $\SI{12.376}{\angstrom}$  \cite{gellerCrystalStructureFerrimagnetism1957}  \\
		$S$ &  14.2 \\
		$4\pi M_s$ & $\SI{1750}{G}$  \cite{tittmannPossibleIdentificationMagnetostatic1973}  \\
		$J$ & $\SI{1.60}{K}$ \cite{kreiselMicroscopicSpinwaveTheory2009} \\
		$\alpha_G$ & $\num{e-4}$ \cite{haertingerSpinPumpingYIG2015} \\
		$\alpha_{IF}$ & $\num{e-2}$ \cite{haertingerSpinPumpingYIG2015} \\
		$\mu_{e}$ & $\SI{8}{\mu V}$  \cite{cornelissenMagnonSpinTransport2016} \\
	\end{tabular}
\end{ruledtabular}
\end{table}

\begin{figure*}
	\centering
	\includegraphics{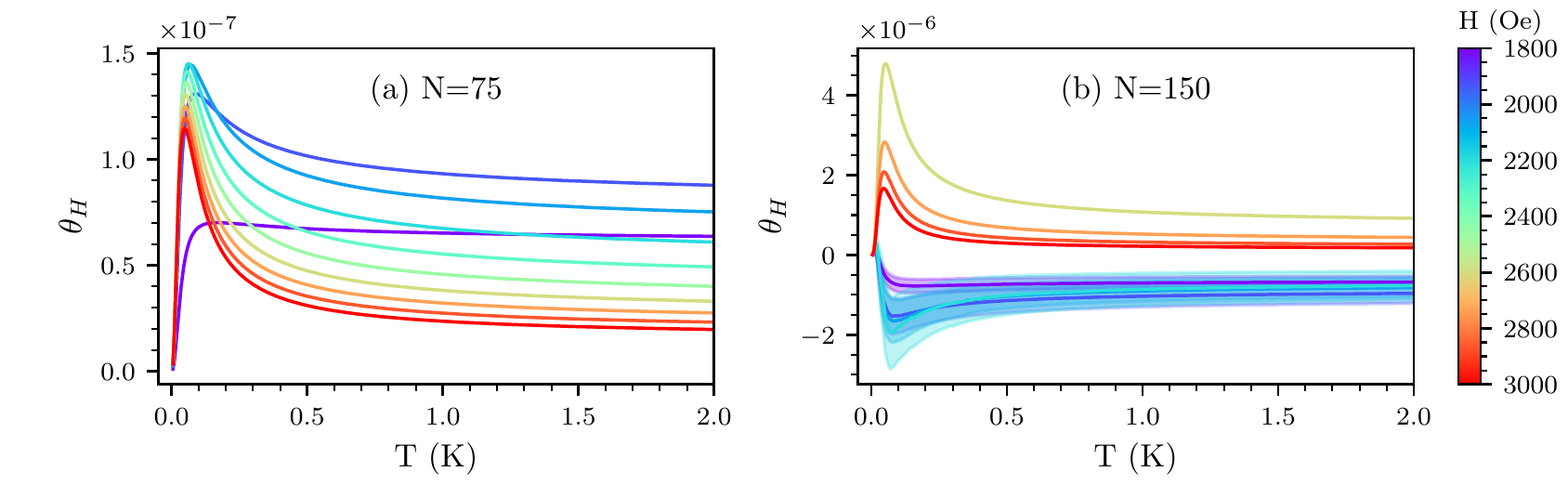}
	\caption{The Hall angle $\Hangle=\sigma_s^H/\sigma_s$ for two different film thicknesses, (a) $N=75$ and (b) $N=150$. The shaded area indicates the error, which results from a slowly converging integral over the Brillouin zone. \label{fig:compare75_150}}
\end{figure*}

\begin{figure*}
	\centering
	\includegraphics{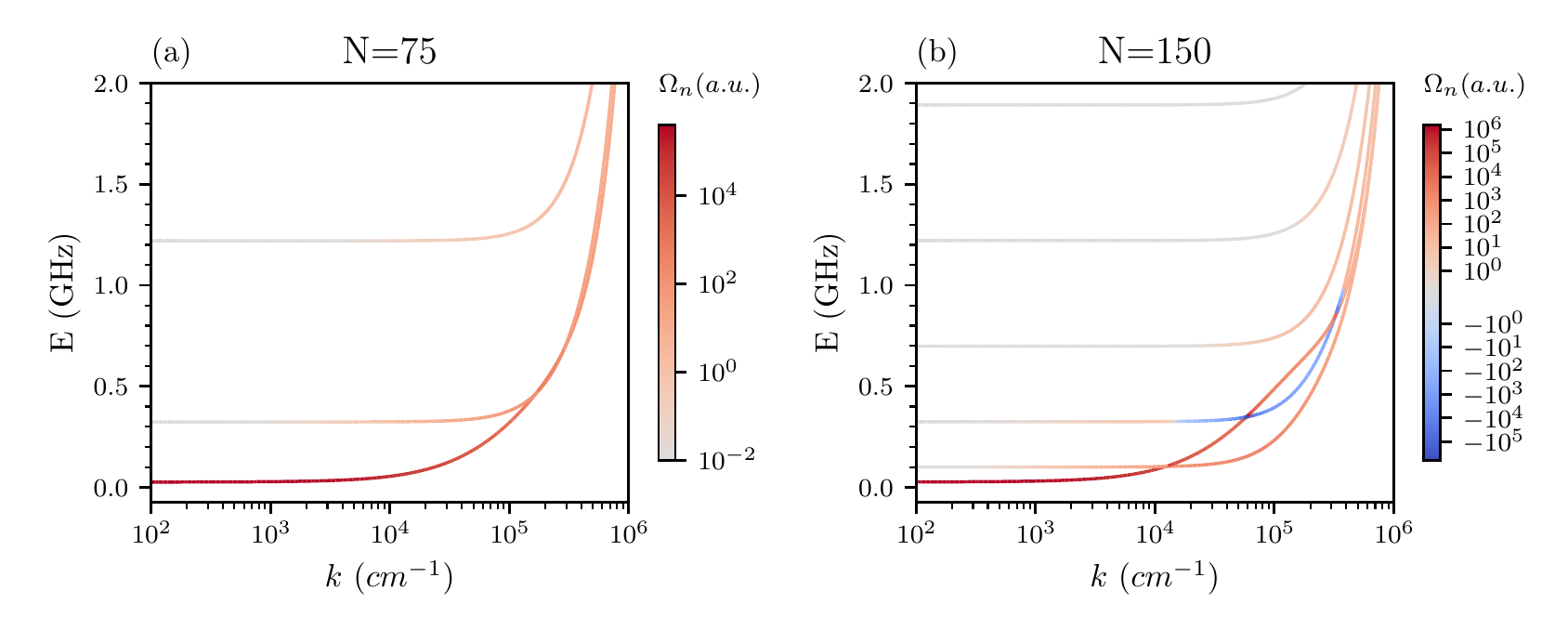}
	\caption{The Berry curvature $\bcurv_{n}^{yz}$ per band for the forward-volume modes of a thin film with (a) $N=75$  and (b) $N=150$ layers, and a magnetic field strength $H=\SI{1800}{Oe}$. Note the more complicated Berry curvature structure for $N=150$, which is not present for the $N=75$ thin film and is due to the band crossings. We also note that the Berry curvature is negative for certain bands for $N=150$, but for none for $N=75$. 
		\label{fig:berry_curvature}}
\end{figure*}

Next, we consider the Hall angle, $\Hangle=\sigma_s^H/\sigma_s$. We compare two films with thicknesses $N=75$ and $N=150$ in Fig.~\ref{fig:compare75_150}. {It is clear that the Hall angle peaks for small temperature, and tends to a lower constant value for higher temperature. }The complete drop-off at $T=0$ is explained by the fact that there are no magnons thermally excited at zero temperature. 
{In order to further explain these results we first need to focus on the Berry curvature for these thin films, since the Berry curvature is directly related to the Hall conductivity in this system.} We therefore show the Berry curvature $\bcurv^{yz}_n$ of the $n$-th band in Fig.~\ref{fig:berry_curvature} for these two films. We can see that the Berry curvature is largest for the lowest band, which we therefore expect to dominate transport. 
Furthermore, in the dipolar regime, at small wavevectors, the Berry curvature is largest. This explains the temperature dependence of $\Hangle$ we observe in Fig.~\ref{fig:compare75_150}. At low temperatures the dipolar magnons dominate transport, and they have a large Berry curvature. Furthermore, the exchange bands naturally have a larger contribution to transport than the dipolar magnons (not shown here). {As the temperature increases, the ratio between the exchange and dipolar magnons shift towards the exchange magnons, increasing the magnon spin conductivity, but not the Hall conductivity.} 

For the film with thickness $N=150$, shown in Fig.~\hyperref[fig:compare75_150]{\ref*{fig:compare75_150}b}, {the Hall angle is negative for low magnetic field}. Here the shaded region indicates the error from integrating the Berry curvature $\Omega_n$ over the Brillouin zone. {The larger errors can be explained from the behaviour of the Berry curvature} close to band crossings, as shown in Fig.~\hyperref[fig:berry_curvature]{\ref*{fig:berry_curvature}b}. The Berry curvatures grows at band crossings---but never diverges, since none of the bands are ever degenerate. This can also be seen from Eq.~\eqref{eq:berryphase}, where it is clear that the Berry connection matrix and therefore the Berry curvature of the band $n$ is inversely proportional to the energy gap. Integrating such a function is numerically very costly, and we only reach the precision as indicated by the shaded region. The avoided band crossings in the dispersion, which lead to an increased Berry curvature, are only present for thicker films ($N\gtrsim 150$).
The results for the Berry curvature can directly be compared to the Berry curvature as obtained by Okamoto and Murakami \cite{okamotoBerryCurvatureMagnons2017}, who showed the same behaviour as we have shown for the $N=150$ film, with an enhanced Berry curvature at the band crossings and a negative Berry curvature for some of the higher bands.

{The negative Hall angle can be explained from the negative Berry curvature}, which is present for $N=150$, but not for $N=75$, as was shown in Fig.~\ref{fig:berry_curvature}. This sign switch of the Hall angle is similar to what was observed by Hirschberger \textit{et al.} \cite{hirschbergerThermalHallEffect2015} in measuring the thermal Hall effect in a Kagome magnet.

\begin{figure}
	\centering
	\includegraphics{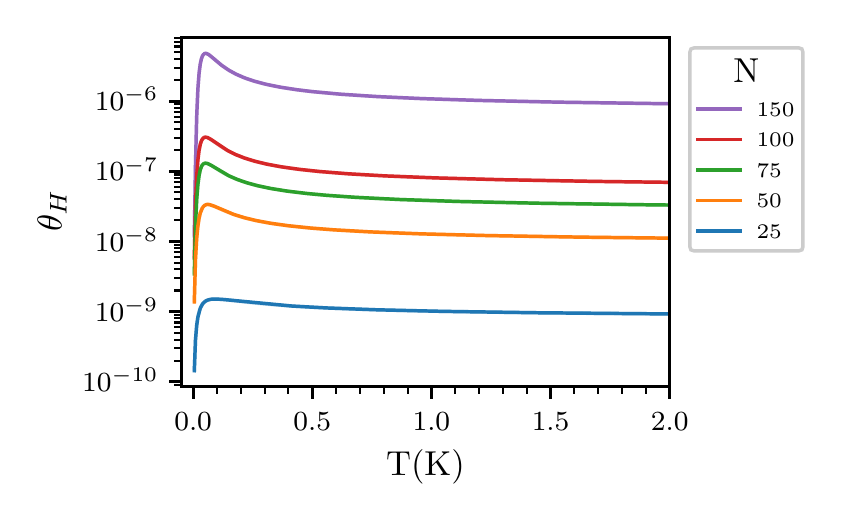}
	\caption{The Hall angle $\Hangle$ for $H=\SI{2600}{Oe}$, as a function of temperature and for varying thicknesses. We were not able to numerically calculate $\Hangle$ for thicker films, so it is not clear if the Hall angle will continue to increase. \label{fig:H2600}}
\end{figure}

For the forward volume modes, the magnetic field acts as a way to introduce a finite energy shift of the bands. {This can be used to explain the behaviour of the spin diffusion length} as shown in Fig.~\ref{fig:diffusionlength}. A higher magnetic field reduces the diffusion length, since by shifting all the bands the magnetic field changes which bands are occupied and therefore contribute. For the Hall angle, $\Hangle$, the magnetic field dependence is more complicated, at least for smaller magnetic fields. As a function of magnetic field strength, the Hall angle rises rapidly, until it peaks for a field of strength $\sim\SI{2400}{Oe}$, after which it drops again. {For higher fields, the magnetic field essentially shifts the ratio between which type of magnons contribute at a given energy: the exchange or the dipolar magnons.} This does not explain the low magnetic field behaviour though, since we expect this behaviour to be (roughly) linear. Further research is needed to understand this in more detail.

Since we have determined that thickness plays a role in the Hall effect of YIG, we also show the results for a fixed magnetic field, with increasing thickness in Fig.~\ref{fig:H2600}. {It can clearly be observed that the Hall angle increases for thicker films.} However, one should be aware that this is still assuming that there is no diffusive transport along the film normal, i.e. the spin diffusion length is larger than the film thickness. The spin diffusion length for YIG thin films at the temperature range considered here has not yet been measured, but for $T=\SI{30}{K}$ it is roughly \SI{5}{\micro m} \cite{cornelissenTemperatureDependenceMagnon2016}, which would make our description valid for thin films up to $N=5000$.

{We have now calculated the transport coefficients $\sigma_s, \sigma_s^H$ and $\Gamma_{s}$.} Not discussed in the main text are the coefficients $A, B$ and $C$ that govern spin injection at the metallic lead interface, which we show in App. \ref{app:spinsink}. Next, we solve the diffusion-relaxation equation, {in order to determine if the magnon Hall effect can be measured electrically.}

\section{Diffusion in the Hall bar}
\label{sec:results}
{Experimentally, the main observable is the difference between the spin currents detected by terminals 2 and 4. }
We define a Hall coefficient as the signal difference between detectors 2 and 4, 
\begin{equation}
	\Delta I = \frac{I_2 - I_4}{I_2 + I_4}.
	\label{eq:deltaI}
\end{equation} 

In order to confirm that a non-zero Hall angle $\Hangle$ results in a finite $\Delta I$ we {numerically solve the diffusion-relaxation equation.}
We choose $M=\SI{8}{\micro m}$, $L_a=\SI{3}{\micro m}$ and $L_b=\SI{0.1}{\micro m}$, which are the same dimensions used by Das \textit{et al.} \cite{dasEfficientInjectionDetection2018} to measure the planar Hall effect in YIG.
The distribution of the chemical potential for a typical system is shown in Fig.~\ref{fig:hall-bar}. The chemical potential diffuses through the film and gets picked up by the three detectors. Note that the difference between the currents picked up by detectors 2 and 4, i.e. $\Delta I$, is too small to be visible on the color scale of Fig.~\ref{fig:hall-bar}. 

\begin{figure}
	\centering
	\includegraphics{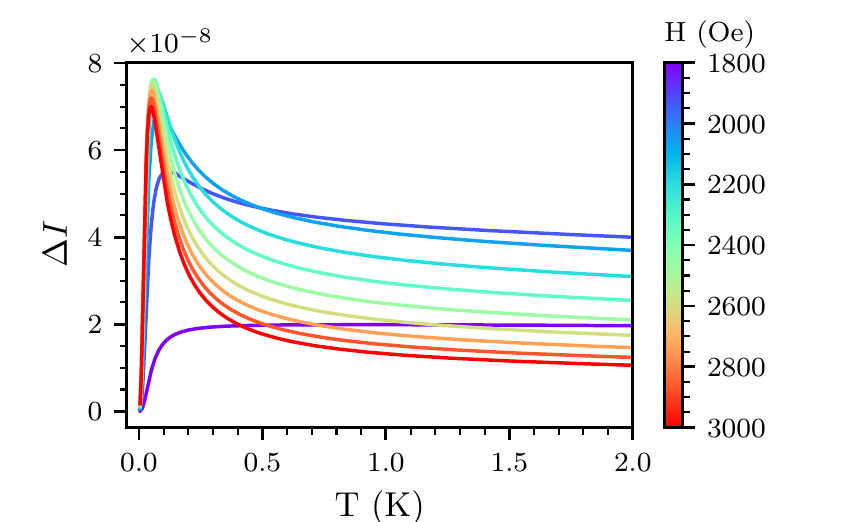}
	\caption{The Hall coefficient $\Delta I$, which follows from the numerical solution to the diffusion-relaxation equation for a Hall bar geometry. The thickness of the film is $N=75$ and this can therefore be directly compared to the Hall angle $\Hangle$ in Fig.~\hyperref[fig:compare75_150]{\ref*{fig:compare75_150}a}. From this comparison it is clear that a Hall response can be measured, and that $\Hangle$ is a direct predictor of $\Delta I$.  \label{fig:hallsignal}}
\end{figure}
{We then calculate the Hall coefficient $\Delta I$ for $N=75$ and show the results in Fig.}\ref{fig:hallsignal}. These results can be compared to the Hall angle, $\Hangle$, in Fig.~\hyperref[fig:compare75_150]{\ref*{fig:compare75_150}a}. From this comparison it is clear that the Hall angle $\Hangle$ is directly related to the Hall coefficient $\Delta I$. We see little to no effect from the magnon relaxation, since the spin diffusion length is much longer than the size of the Hall bar. Most importantly, there are no (large) corrections from interface effects due to the electrical injection and detection. This is also the case for different thicknesses. {We therefore conclude that the magnon Hall effect can in principle be measured electrically in a Hall bar geometry. }

\section{Conclusion and Discussion}
\label{sec:conclusions}

We have derived and calculated the anomalous Hall conductivity for magnons in a thin film of YIG, using a microscopic model. Furthermore, we have shown that a non-zero anomalous Hall conductivity results in a measurable signal in a Hall bar setup and can be measured electrically. The magnon Hall effect has previously only been measured thermally in materials with a Dzyaloshinskii-Moriya spin-orbit interaction \cite{onoseObservationMagnonHall2010}, but with a Hall bar setup as discussed here this magnon Hall effect could also be measured electrically in YIG.

Using realistic parameters we have calculated the size of the expected Hall angle, and its dependency on temperature and magnetic field. Moreover, we have shown that for thicker films of YIG, there is a sign change in the Hall angle as a function of the magnetic field, which would be a strong experimental indicator of the magnon Hall effect.

The presented method can be applied to any bosonic system with anomalous coefficients to determine anomalous transport properties. In fact, the physical origin of the anomalous transport properties discussed here are the dipole-dipole interactions, which are universally present in any magnetic system. As such, this method can be applied to a wide range of magnetic materials.

In order to measure this effect it is possible to use the fact that the sign of the Hall angle switches as the field is reversed. Therefore, by comparing measurements with opposite field, the anomalous contributions can be isolated. This is especially useful since the spin diffusion and relaxation means that the distance between the injector at lead 1 and the detectors at leads 2 and 4 is critical. 

As was shown by Takahashi and Nagaosa \cite{takahashiBerryCurvatureMagnonPhonon2016} and Okamoto \textit{et al.} \cite{okamotoBerryCurvatureMagnetoelastic2020}, for magnetoelastic waves the Berry curvature is enhanced at the crossing of the magnon and phonon branches. This could therefore serve to further enhance the magnon Hall effect discussed here. The inclusion of magnon-phonon coupling on our formalism is left for future work.

\begin{acknowledgments}
	R.D. is member of the D-ITP consortium, a program of the Dutch Organization for Scientific Research (NWO) that is funded by the Dutch Ministry of Education, Culture and Science (OCW). This project has received funding from the European Research Council (ERC) under the European Union’s Horizon 2020 research and innovation programme (grant agreement No. 725509). This work is part of the research programme of the Foundation for Fundamental Research on Matter (FOM), which is part of the Netherlands Organization for Scientific Research (NWO). We thank Timo Kuschel for discussions and Ruben Meijs for doing his thesis work on this subject. 
\end{acknowledgments}

\appendix
\section{Quantum Kinetic Equations}
\label{app:qke}
In this appendix we derive the equation of motion for the spin density of a bosonic Hamiltonian. 
We start from the quantum kinetic equations:
\begin{align}
	\left(\hat{\epsilon}-\nu\hat{H}\right)\hat{G}^{K} & =\nu\hat{\Sigma}^{K}\hat{G}^{A}+\nu\hat{\Sigma}^{R}\hat{G}^{K};
	\label{eq:dyson1}\\
	\hat{G}^{K}\left(\hat{\epsilon}-\hat{H}\nu\right) & =\hat{G}^{R}\hat{\Sigma}^{K}\nu+\hat{G}^{K}\hat{\Sigma}^{A}\nu \label{eq:dyson2},
\end{align}
where hats indicate matrices in space and time, $\hat{\epsilon}=\delta\left(\boldsymbol{r}-\boldsymbol{r}'\right)\delta\left(t-t'\right)i\hbar\partial_{t'}$, $\hat{\Sigma}^{R/A/K}$ are the retarded/advanced/Keldysh self-energies, $\hat{G}^{R/A/K}$ are the retarded/advanced/Keldysh Green's functions and $\nu=\diag\left[1,...,1,-1,...,-1\right]$. We apply a Wigner transformation, defined as
\begin{multline*}
	A\left(\boldsymbol{r},t;\boldsymbol{p},\varepsilon\right)=\int dr'\int dt' \\
	\hat{A}\left(\boldsymbol{r}+\frac{\boldsymbol{r}'}{2},t+\frac{t'}{2};\boldsymbol{r}-\frac{\boldsymbol{r}'}{2},t-\frac{t'}{2}\right)e^{-i\left(\boldsymbol{k}\cdot\boldsymbol{r}'-\omega t'\right)}
\end{multline*}
and expand up to first order in $\hbar$, such that we have (suppressing all labels from here on)
\begin{align}
	\bigg(\varepsilon-\nu H-\nu\Sigma^{R}+\frac{i\hbar}{2}\partial_{t}\qquad& \nonumber \\
    + \frac{i}{2}\nu\left(\nabla_{p}H\right)\cdot\nabla_{r}\bigg)G^{K} &=\nu\Sigma^{K}G^{A}; \\
	G^{K}\bigg(\varepsilon-H\nu-\Sigma^{A}\nu-\frac{i\hbar}{2}\overleftarrow{\partial}_{t} \qquad\nonumber& \\
	-\frac{i}{2}\overleftarrow{\nabla}_{r}\cdot\left(\nabla_{p}H\right)\nu\bigg) &= G^{R} \Sigma^{K}\nu,
\end{align}
where we assume that the Hamiltonian does not depend explicitly on position or time, i.e. $H\left(\boldsymbol{r},t;\boldsymbol{k},\omega\right)=H\left(\boldsymbol{k}\right)$ and have used arrows to indicate to which function the derivative applies, if there are ambiguities. 

Furthermore, we define a covariant derivative as 
\begin{equation}	D_{k_{\alpha}}\mathcal{E}\equiv\mathcal{T}^{\dagger}\left(\partial_{k_{\alpha}}\mathcal{H}\right)\mathcal{T}=\partial_{k_{\alpha}}\mathcal{E}+i\mathcal{E}A^{\alpha}-i\nu\mathcal{E}\nu A^{\alpha}.
	\label{eq:covariant}
\end{equation}
We introduce the transformed Green's functions $g^{R/A/K}	=\mathcal{T}^{-1}G^{R/A/K}\left(\mathcal{T}^{\dagger}\right)^{-1}$ and self-energies 
$\sigma^{R/A/K}	=\mathcal{T}^{\dagger}\Sigma^{R/A/K}\mathcal{T}$ and assume damped quasiparticles in (local) thermal equilibrium, such that 

\begin{align}
	\sigma^{R/A}\left(\boldsymbol{k},\omega\right) & =\mp i\nu\left[ \Gamma^{\mm}\left(\boldsymbol{k},\omega\right) + \Gamma^{\mr}\left(\boldsymbol{k},\omega\right) \right];  \\
	\sigma^{K}\left(\boldsymbol{r};\boldsymbol{k},\omega\right) & =-2i\Gamma^{\mm}\left(\boldsymbol{k},\omega\right)F_n\left(\boldsymbol{r},\omega\right)\nonumber \\
	& \hspace{3em} - 2i\Gamma^{\mr}\left(\boldsymbol{k},\omega\right)F_n^{\mu_m=0}\left(\boldsymbol{r},\omega\right),
\end{align}
where 
\begin{equation*}
	\Gamma_{mn}^{\eta}\left(\kv,\omega\right)=\delta_{mn}\times
	\begin{cases}
		\gamma^{\eta}\left(\boldsymbol{k},\omega\right) & 1\leq n\leq N,\\
		\gamma^{\eta}\left(-\boldsymbol{k},-\omega\right) & N+1\leq n\leq2N,
	\end{cases}
\end{equation*}
with $\eta \in \{\mr,\mm\}$ representing the magnon relaxation processes (which do not conserve spin) and magnon-magnon interactions (which conserve spin) respectively. The distribution function is defined as
\begin{equation*}
	F_{mn}\left(\boldsymbol{r},\omega\right)=\delta_{mn}\times\begin{cases}
		f_{B}\left(\boldsymbol{r},\omega\right) & 1\leq n\leq N,\\
		f_{B}\left(\boldsymbol{r},-\omega\right) & N+1\leq n\leq2N,
	\end{cases}
\end{equation*}
where $f_{B}=\coth\left(\frac{\hbar\omega-\mu_m}{2k_BT}\right)$ is the symmetrized Bose-Einstein distribution. The distribution function $F_n^{\mu_m=0}\left(\boldsymbol{r},\omega\right)$ describes the relaxation of magnons to the lattice. For brevity, we write $\Gamma_{n}\left(\boldsymbol{k},\omega\right)=\Gamma_{n}^{\mr}\left(\boldsymbol{k},\omega\right) + \Gamma_{n}^{\mm}\left(\boldsymbol{k},\omega\right)$. 

The retarded and advanced Green's functions are then given by 
\begin{equation}
	g^{R/A}=\frac{\delta_{nm}}{\nu_{n}\left(\hbar\omega\pm i\Gamma_n\right)-\mathcal{E}_{n}}.
\end{equation}
For the Keldysh Green function we first solve the diagonal component of the distribution function, $f_{n}\equiv\frac{i\hbar}{2}g_{nn}^{K}$, using the difference between Eqs.~(\ref{eq:dyson1}) and (\ref{eq:dyson2}), such that 
\begin{widetext}

\begin{equation}
	\partial_{t}f_{n}+\sum_{\alpha}\partial_{r_{\alpha}}j_n^{\alpha}
	=
	-2\frac{\Gamma^{\mr}_n}{\hbar}\left[ f_n -  \frac{\hbar\Gamma_{n}}{\left(\hbar\omega-\nu_{n}\mathcal{E}_{n}\right)^{2}+\Gamma_{n}^{2}}F^{\mu_m=0}_{n} \right] 
	-2\frac{\Gamma^{\mm}_{n}}{\hbar}\left[ f_n -  \frac{\hbar\Gamma_{n}}{\left(\hbar\omega-\nu_{n}\mathcal{E}_{n}\right)^{2}+\Gamma_{n}^{2}}F_{n} \right],
	\label{eq:difference}
\end{equation}

where the current density 
\begin{equation}
	j_n^{\alpha}=\frac{\nu_{n}}{\hbar}\left(\partial_{k_{\alpha}}\mathcal{E}_{n}\right)f_{n}  +\nu_{n}\frac{i}{4}\sum_{m\neq n}\left[\left(D_{k_{\alpha}}\mathcal{E}\right)_{nm}g_{mn}^{K}+g_{nm}^{K}\left(D_{k_{\alpha}}\mathcal{E}\right)_{mn}\right],
	\label{eq:currentdensity}
\end{equation}
has contributions from the off-diagonal components. 

We now assume that there is local thermal equilibrium, and thus that the local distribution function $f_n$ can be described by small corrections $\delta f_n$ on top of the thermal equilibrium. This is possible because the spin conserving processes (represented by $\Gamma^{\mm}$) are much faster than the non-spin-conserving processes (represented by $\Gamma^{\mr}$). Thus, we disregard the $F_n^{\mu_m=0}$ term in Eq.~(\ref{eq:difference}) and make the ansatz 
\begin{equation}
	f_{n}=\frac{\hbar\Gamma_{n}}{\left(\hbar\omega-\nu_{n}\mathcal{E}_{n}\right)^{2}+\Gamma_{n}^{2}}F_{n} +\delta f_n,
	\label{eq:ansatz}
\end{equation}
where $\delta f_n$ is at least one order higher in gradients. In a steady state (such that $\partial_t f_n=0$) we further note that from Eq.~(\ref{eq:difference}) it is clear that 
\begin{equation}
	\sum_{\alpha}\partial_{r_{\alpha}}j_n^{\alpha} = 
	-2\frac{\Gamma_{n}}{\hbar}\left( \delta f_n + \frac{\Gamma^{\mr}_{n}}{\left(\hbar\omega-\nu_{n}\mathcal{E}_{n}\right)^{2}+\Gamma_{n}^{2}}\left(F_{n} - F_{n}^{\mu_m=0} \right)  \right)
	.
\end{equation}
This can then be solved up to first order in gradients by inserting the ansatz, Eq.~(\ref{eq:ansatz}), into the current density, Eq.~(\ref{eq:currentdensity}), and using the fact that $g^K_{nm}$ is one order higher in gradients and can thus be discarded. Then we find 
\begin{equation}
	\delta f_n = 
	-\frac{\nu_{n}\hbar}{2}\left(\partial_{k_{\alpha}}\mathcal{E}_{n}\right) \frac{1}{\left(\hbar\omega-\nu_{n}\mathcal{E}_{n}\right)^{2}+\Gamma_{n}^{2}} \left(\partial_{r_{\alpha}}F_{n}\right) \\
	 - \frac{\hbar\Gamma^{\mr}_{n}}{\left(\hbar\omega-\nu_{n}\mathcal{E}_{n}\right)^{2}+\Gamma_{n}^{2}} \left( F_n - F_n^{\mu_m=0} \right).
\end{equation}
 
 In order to find $g_{nm}^K$ we consider the sum of Eqs.~(\ref{eq:dyson1}) and (\ref{eq:dyson2}) and find for $m \neq n$ that
\begin{equation}
	\left[2\hbar\omega-\nu_{n}\mathcal{E}_{n}-\nu_{m}\mathcal{E}_{m}+i\left(\Gamma_{n}-\Gamma_{m}\right)\right]g_{nm}^{K}
	=
	-
	\frac{i}{2}\sum_{\alpha}\sum_{l}\left[\nu_{n}\left(D_{k_{\alpha}}\mathcal{E}\right)_{nl}\left(\partial_{r_{\alpha}}g_{lm}^{K}\right)-\nu_{m}\left(\partial_{r_{\alpha}}g_{nl}^{K}\right)\left(D_{k_{\alpha}}\mathcal{E}\right)_{lm}\right].
	\label{eq:gmn}
\end{equation}
It is convenient to proceed in the quasiparticle limit ($\lim_{\Gamma_n \rightarrow 0^+}$), where
\begin{equation}
	\lim_{\Gamma \rightarrow 0^+} f_{n}=\pi\delta\left(\omega-\nu_{n}\mathcal{E}_{n}/\hbar\right)F_{n}\left(\omega\right)+\delta f_n.
\end{equation}
We now use the fact that $g^K_{nm}$ is one order higher in gradients than $f_n$, and as such can write

	\begin{equation}
		g_{nm}^{K}=\frac{1}{\hbar}\frac{i\pi}{\nu_{m}\mathcal{E}_{n}-\nu_{n}\mathcal{E}_{m}}\sum_{\alpha} \partial_{r_{\alpha}}\left(D_{k_{\alpha}}\mathcal{E}\right)_{nm} \Big[\nu_{m}\delta\left(\omega-\nu_{n}\mathcal{E}_{n}/\hbar\right)F_{n}\left(\omega\right)+\nu_{n}\delta\left(\omega-\nu_{m}\mathcal{E}_{m}/\hbar\right)F_{m}\left(\omega\right)\Big];\quad(m\neq n), 
	\end{equation}
where we have used the diagonal components $f_n$ to rewrite Eq.~(\ref{eq:gmn}), only keeping terms up to first order in gradients. 
Using the definition of the covariant derivative in Eq.~(\ref{eq:covariant}) we now write the current as 
	\begin{multline}
		j_{n}^{\alpha} = \nu_{n}\left(\partial_{k_{\alpha}}\mathcal{E}_{n}\right)\pi\delta\left(\omega-\nu_{n}\mathcal{E}_{n}/\hbar\right)\left[F_{n} -\frac{\Gamma_n^{\mr}}{\Gamma_n}\left( F_{n} - F_{n}^{\mu_m=0} \right) \right]
		- \frac{1}{2\Gamma_{n}\hbar}\sum_{\beta}\left(\partial_{k_{\alpha}}\mathcal{E}_{n}\right)\left(\partial_{k_{\beta}}\mathcal{E}_{n}\right)\pi\delta\left(\omega-\nu_{n}\mathcal{E}_{n}/\hbar\right)\partial_{r_{\beta}}F_{n}
		\\
		 +\frac{i}{4\hbar}\sum_{m\neq n}\sum_{\beta}\left(\nu_{n}\nu_{m}\mathcal{E}_{m}-\mathcal{E}_{n}\right)\left(A_{mn}^{\alpha}A_{nm}^{\beta}-A_{mn}^{\beta}A_{nm}^{\alpha}\right)
		 \partial_{r_{\beta}}\left[\nu_{n}\pi\hbar\delta\left(\omega-\nu_{m}\mathcal{E}_{m}/\hbar\right)F_{m}+\nu_{m}\pi\hbar\delta\left(\omega-\nu_{n}\mathcal{E}_{n}/\hbar\right)F_{n}\right],
		 \label{eq:current_full}
	\end{multline}
such that we now have a full description of the equation of motion, Eq.~(\ref{eq:difference}) for the distribution function of the magnons. Note that the first term in Eq.~(\ref{eq:current_full}) will be zero if integrated over, due to inversion symmetry.

We continue with the spin density, which is defined as 
\begin{align}
	s^z\left(\boldsymbol{r},t\right)&=-\frac{i\hbar}{4}\Tr \left[\hat{G}^{K}\right],
	\nonumber \\
	&=-\frac{i\hbar}{4}\int\frac{d^{d}k}{\left(2\pi\right)^{d}}\int\frac{d\omega}{2\pi}\Tr\left[\mathcal{T}^{\dagger}\mathcal{T}g^{K}\right],
\end{align}
such that 
\begin{equation}
	\partial_{t}s^z\left(\boldsymbol{r},t\right) = \frac{1}{2}\int\frac{d^{d}k}{\left(2\pi\right)^{d}}\int\frac{d\omega}{2\pi}\sum_{n} \left(\mathcal{T}^{\dagger}\mathcal{T}\right)_{nn}
	\left[
	\sum_{\alpha}\partial_{r_{\alpha}}j_{n}^{\alpha} +
	2\frac{\Gamma^{\mr}}{\hbar}\left( f_n -  \frac{\hbar\Gamma_{n}}{\left(\hbar\omega-\nu_{n}\mathcal{E}_{n}\right)^{2}+\Gamma_{n}^{2}}F^{\mu_m=0}_{n} \right) 	
	\right],
	\label{eq:spindensity}
\end{equation}
where we have only kept terms up to first order in gradients. Since the processes described by $\Gamma^{\mm}$ always conserve spin and because we assume them to approximately conserve momentum, we furthermore disregard all terms related to $\Gamma^{\mm}$, such that $\Gamma_n=\Gamma_n^{\mr}$. Its inclusion up to this point was however necessary, since without it a local thermal equilibrium cannot be properly defined and a current density cannot be expressed in terms of the magnon chemical potential.

\section{Coefficients}
\label{app:coefficents}
From here on, we assume Gilbert damping for the magnon relaxation process, such that $\gamma^{\mr}\left(\boldsymbol{k},\omega\right)=2\alpha_{G} \hbar\omega$ \cite{benderDynamicPhaseDiagram2014}, where $\alpha_G$ is the bulk Gilbert damping parameter. With the generic equation of motion, Eq.~(\ref{eq:spindensity}), we now derive the equation of motion up to linear order in the magnon chemical potential, giving
\begin{equation}
	\partial_t s^z\left(\boldsymbol{r},t\right) + \sum_\alpha \partial_{r_\alpha}J_{s}^{\alpha} = \Gamma_s \mu_m,
\end{equation}
where $J_{s}^{\alpha} = \sigma^{\alpha\alpha}\partial_{r_{\alpha}}\mu_m+\sum_{\beta}\sigma^{\alpha\beta}\partial_{r_{\beta}}\mu_m$, with 

	\begin{align}
		\sigma^{\alpha\alpha} &= -\frac{1}{32\hbar\alpha_{G} k_BT}\int\frac{d^{2}k}{\left(2\pi\right)^{2}}\sum_{n}\left(\mathcal{T}^{\dagger}\mathcal{T}\right)_{nn}\frac{\left(\partial_{k_{\alpha}}\mathcal{E}_{n}\right)^{2}}{\mathcal{E}_{n}}\csch\left[\frac{\mathcal{E}_{n}}{2k_BT}\right]^{2};  \\
		\sigma^{\alpha\beta}&=\frac{1}{32k_BT\hbar}\int\frac{d^{2}k}{\left(2\pi\right)^{2}}\sum_{n,m,m\neq n}\left(\nu_{n}\left(\mathcal{T}^{\dagger}\mathcal{T}\right)_{nn}+\nu_{m}\left(\mathcal{T}^{\dagger}\mathcal{T}\right)_{mm}\right)
		\left(\nu_{n}\nu_{m}\mathcal{E}_{m}-\mathcal{E}_{n}\right)
		\Omega^{\alpha\beta}_{m}
		\csch\left[\frac{\mathcal{E}_{n}}{2k_BT}\right]^{2};\\
		\Gamma_{s}&=-\frac{1}{2k_BT}\int\frac{d^{2}k}{\left(2\pi\right)^{2}}\int\frac{d\omega}{2\pi}\sum_n \left(\mathcal{T}^{\dagger}\mathcal{T}\right)_{nn}\frac{\left(2\alpha_{G}\hbar\omega\right)^2}{\left(\hbar\omega-\nu_{n}\mathcal{E}_{n}\right)^{2}+\left(2\alpha_{G}\hbar\omega\right)^2}\csch\left[\frac{\nu_n\hbar\omega}{2k_BT}\right]^{2}.
	\end{align}

\end{widetext}

Here we have disregarded the $\Gamma^{\mm}$ term, since magnon-magnon scattering preserves momentum and should therefore not contribute to the magnon spin conductivity $\sigma^{\alpha\alpha}$. We then have $\sigma_s=\sigma^{xx}=\sigma^{yy}$ and $\sigma_s^H=\sigma^{xy}$, since the system is rotationally invariant.

In order to calculate these coeffiencts we diagonalize the Hamiltonian $H$ with a paraunitary matrix $\mathcal{T}$, which also gives the energies $\mathcal{E}$. Moreover, we construct $\partial_{k_\alpha}H$, such that we calculate the Berry phase and subsequently the Berry curvature using Eq.~(\ref{eq:berryphase}). These terms are further shown in Appendix~\ref{app:hamiltonian}. We can then integrate the coefficients $\sigma_s^H,\sigma_s$ and $\Gamma_s$ over the entire Brillouin zone, where we use the translation invariance to employ the one-dimensional Gauss–Kronrod quadrature formula, which also gives an error estimate. These results are shown in Sec.~\ref{sec:coefficients}.
\section{Metallic lead}
\label{app:spinsink}
We now consider how the equation of motion for the spin density has to be modified if a metallic lead is interfaced to the ferromagnet.
Attaching a metallic lead, the self-energies are modified such that ${\Sigma}^{R/A/K}={\Sigma}_{\text{bulk}}^{R/A/K}+{\Sigma}_{\text{IF}}^{R/A/K}$, with
\begin{align}
	\Sigma_{\text{IF}}^{R/A}\left(\bm{r},t;\bm{k},\omega\right)	&=\mp i \alphaIF \left(\hbar\omega-\nu\mu_e\right), \\
	\Sigma_{\text{IF}}^{K}\left(\bm{r},t;\bm{k},\omega\right)	&=2\Sigma_{IF}^{R} F_{e}\left(\omega\right),
\end{align}
where 
\begin{equation}
	F_{e}\left(\boldsymbol{r},\omega\right)=\delta_{nm}\times\begin{cases}
		\coth\left[\frac{\hbar\omega-\mu_{e}}{2k_BT}\right] & 1\leq n\leq N,\\
		\coth\left[\frac{-\hbar\omega-\mu_{e}}{2k_BT}\right] & N+1\leq n\leq2N,
	\end{cases}
\end{equation}
and $\alphaIF$ is the interfacial Gilbert damping. 
The equation of motion for the spin density, Eq.~(\ref{eq:spindensity}), is then modified to $\partial_{t}s^z+\nabla\cdot \bm{J}^{s}=\Gamma_{s}+\Gamma_{s}^{\text{IF}}$, where
\begin{widetext}
\begin{equation}
\Gamma_{s}^{\text{IF}}	=-\frac{1}{4}\int\frac{d^{d}k}{\left(2\pi\right)^{d}}\int\frac{d\omega}{2\pi}Tr\left[\mathcal{T}^{\dagger}\nu\Sigma_{\text{IF}}^{K}\mathcal{T}g^{A}-\mathcal{T}^{\dagger}\nu\Sigma_{\text{IF}}^{K}\mathcal{T}g^{R}+\mathcal{T}^{\dagger}\nu\Sigma_{\text{IF}}^{R}\mathcal{T}g^{K}-\mathcal{T}^{\dagger}\nu\Sigma_{\text{IF}}^{A}\mathcal{T}g^{K}\right].
\end{equation}
Noting that, up to lowest order in the interfacial coupling, the Green's functions $g^{R/A/K}$ are unchanged by the interfacial self-energies, we can further write this as (in the quasiparticle limit)

\begin{equation}
	\Gamma_{s}^{\text{IF}}
	=
	\frac{\alphaIF}{2\hbar}
	\int\frac{d^{d}k}{\left(2\pi\right)^{d}}
	\Tr
	\left[\left(\mathcal{T}^{\dagger}\left(\mathcal{E}-\mu_{e}\right)\mathcal{T}
	F\left(\nu\mathcal{E}\right)
	-\mathcal{T}^{\dagger}\nu\left(\mathcal{E}-\mu_{e}\right)F_{e}\left(\nu\mathcal{E}\right)\mathcal{T}\nu\right)\right].
\end{equation}
We again keep only terms linear in $\mu_m$ and $\mu_e$, such that we can write $\Gamma_{s}^{\text{IF}}=A\mu_m + B\mu_e + C$, with

	\begin{align}
		A & =\frac{\alphaIF}{4\hbar k_BT} \int\frac{d^{2}k}{\left(2\pi\right)^{2}}\Tr\left[\mathcal{T}^{\dagger}\mathcal{E}\mathcal{T}\csch\left[\frac{\mathcal{E}}{2k_BT}\right]^{2}\right];\\
		B & =-\frac{\alphaIF}{2\hbar} \int\frac{d^{2}k}{\left(2\pi\right)^{2}}\Tr\left[\mathcal{T}^{\dagger}\mathcal{T}\coth\left[\frac{\mathcal{E}}{2k_BT}\right]-\frac{1}{2k_BT}\mathcal{E}\csch\left[\frac{\mathcal{E}}{2k_BT}\right]^{2}+\coth\left[\frac{\mathcal{E}}{2k_BT}\right]\right];\\
		C & =\frac{\alphaIF}{2\hbar} \int\frac{d^{2}k}{\left(2\pi\right)^{2}}\Tr\left[\left(\mathcal{T}^{\dagger}\mathcal{E}\mathcal{T}-\mathcal{E}\right)\coth\left[\frac{\mathcal{E}}{2k_BT}\right]\right].
	\end{align}
\end{widetext}

\section{Boundary conditions}
\label{app:bc}

With the equation of motion for the spin density completely determined, we can now consider the boundary condition for the spin density in the Hall bar geometry.
For the metal strips we assume a thin strip, where $L_a\ll L_b$ and the long side $L_b$ interfaces the Hall bar, as shown in Fig.~\ref{fig:hall-bar}. Then the detector can be described by Eq.~(\ref{eq:spin-sink-diffusion}), with the boundary condition that the current at its interface with the main region is continuous. Thus we have, 
\begin{equation}
	\int_{\partial S_{i}}\dd{s}\bm{J}_s\cdot\hat{\boldsymbol{n}} = \int_{S_{i}}\dd{S} \left[(\Gamma_s+A)\mu_m+B\mu_e + C \right],
\end{equation}
where $S_i$ is the area of detector $i$. Note thate for the detectors $\mu_e=0$. We now Taylor expand the chemical potential in the detector strip perpendicular to the interface, and integrate over the short side of the strip, keeping only terms linear in $L_a$, which gives the boundary condition
\begin{equation}
	\int_{\partial S_{i}}\dd{s} \bm{J}_s\cdot\hat{\boldsymbol{n}} 
	=
	L_{a}\int_{\partial S_{i}}\dd{s}\left[ \left(\Gamma_s+A\right)\mu_m+B\mu_e+C \right], 
\end{equation}
where we have required that $\bm{J}_s\cdot \hat{\bm{n}}=0$ at the other three sides of the detector. The boundary condition can now be identified as 
\begin{equation}
	\bm{J}_s \cdot \hat{\bm{n}} = J_s^{\text{int}}(\mu_m),
\end{equation}
where 
\begin{equation}
	J_s^{\text{int}}(\mu_m)=L_a\left[\left(\Gamma_s+A\right)\mu_m+B\mu_e+C\right].
\end{equation}

\section{Hamiltonian}
\label{app:hamiltonian}

In order to determine the dynamics of the magnons in the YIG, we describe this system using the Heisenberg spin Hamiltonian \cite{cherepanovSagaYIGSpectra1993a}
\begin{multline}
	\mathcal{H}=
	-\frac{1}{2}\sum_{ij}J_{ij}\bm{S}_{i} \cdot \bm{S}_{j}-\mu\bm{H}_{e}\cdot\sum_{i}\bm{S}_{i}
	\\ -\frac{1}{2}\sum_{ij,i\neq j}\frac{\mu^2}{|\bm{R}_{ij}|^{3}}\left[3\left(\bm{S}_{i}\cdot\hat{\bm{R}}_{ij}\right)\left(\bm{S}_{j}\cdot\hat{\bm{R}}_{ij}\right)-\bm{S}_{i}\cdot\bm{S}_{j}\right],
	\label{eq:spinhamiltonian}
\end{multline}
where the sums are over the lattice sites $\bm{R}_{i}$, with $\bm{R}_{ij}=\bm{R}_i-\bm{R}_j$ and $\hat{\bm{R}}_{ij} = \bm{R}_{ij} / |\bm{R}_{ij}|$. We only consider nearest neighbour exchange interactions, so $J_{ij}=J$ for nearest neighbours and 0 otherwise. Here $\mu = 2\mu_B$ is the magnetic moment of the spins, with $\mu_B=e\hbar/(2m_ec)$ the Bohr magneton. $\bm{H}_{e}$ is the external magnetic field, which we take strong enough to fully saturate the ferromagnet. 

We apply the Holstein Primakoff transformation up to quadratic order,
\begin{equation}
	S^+_i = \sqrt{2S}b_i;\quad S^-_i = \sqrt{2S}b_i^\dagger;\quad S^z_i = S-b_i^\dagger b_i
\end{equation} 
to the Heisenberg spin Hamiltonian, Eq.~(\ref{eq:spinhamiltonian}), and apply the Fourier transformation in the $xy$-plane, introducing $\kv=(k_x, k_y)$. The coordinate system used is summarized in Fig.~\ref{fig:geometry} in the main text. We can now write the quadratic part of the Hamiltonian in the basis $(
b_{\kv}( z_1),...,b_{\kv}( z_N), b_{-\kv}^\dagger( z_1),...,b^\dagger_{-\kv}( z_N)
)^T$ as 
\begin{equation}
	\mathcal{H}_{\kv} = \begin{pmatrix}
		\bm{A}_{\kv} & \bm{B}_{\kv}\\
		\bm{B}_{\kv}^{\dagger} & \bm{A}_{\kv}
	\end{pmatrix},
	\label{eq:d-matrix}
\end{equation}
where the amplitude factors are
\begin{align}
	A_{\kv}(z_{ij}) & =\sum_{\rv_{ij}}e^{-i\kv\cdot\rv}A(z_{i}-z_{j},\rv), \nonumber \\
	& =\delta_{ij}\left[h+S\sum_{n}D_{0}^{zz}(z_{in})\right] \nonumber
	\\
	&\quad-\frac{S}{2}\left[D_{\kv}^{yy}(z_{ij})+ D_{\kv}^{xx}(z_{ij})\right] + SJ_{\kv}(z_{ij}),\\
	B_{\kv}(z_{ij}) & =\sum_{\rv_{ij}}e^{-i\kv\cdot\rv}B(z_{i}-z_{j},\rv), \nonumber \\
	& =-\frac{S}{2}\left[D_{\kv}^{xx}(z_{ij}) -D_{\kv}^{yy}(z_{ij}) 
	+ i D_{\kv}^{xy}(z_{ij})\right],
\end{align}
where
\begin{multline}
	J_{\kv}(z_{ij}) = J[\delta_{ij} ( 6 - \delta_{j1} -\delta_{jN} \\ 
	-2\cos(k_xa)- 	2\cos(k_ya) ) - \delta_{ij+1} - \delta_{ij-1} ],
\end{multline}
$\rv_{ij} = (x_{ij}, y_{ij})$ and $D_{\kv}^{\alpha\beta}(z_{ij})$ describes the dipole-dipole interaction. 

For the Berry curvature we need to calculate $\partial_{k_\alpha} \mathcal{H}_{\kv}$, where $\alpha\in (x,y)$. This is given by
\begin{equation}
	\partial_{k_\alpha}\mathcal{H}_{\kv} = \begin{pmatrix}
		\partial_{k_\alpha}\bm{A}_{\kv} & \partial_{k_\alpha}\bm{B}_{\kv}\\
		\partial_{k_\alpha}\bm{B}_{\kv}^{\dagger} & \partial_{k_\alpha}\bm{A}_{\kv}
	\end{pmatrix},
\end{equation}
where
\begin{align}
	\partial_{k_\alpha} A_{\kv}(z_{ij}) &  = -\frac{S}{2}\left[\partial_{k_\alpha}D_{\kv}^{yy}(z_{ij})+ \partial_{k_\alpha}D_{\kv}^{xx}(z_{ij})\right] \nonumber \\
	&\hspace{6em} + 2SJa\sin(k_\gamma a),\\
	\partial_{k_\alpha} B_{\kv}(z_{ij}) & =-\frac{S}{2}[\partial_{k_\alpha}D_{\kv}^{xx}(z_{ij}) -\partial_{k_\alpha}D_{\kv}^{yy}(z_{ij}) \nonumber  \\
	&\hspace{6em} + i \partial_{k_\alpha}D_{\kv}^{xy}(z_{ij})],
\end{align}

\begin{widetext}
For the dipolar sums we apply the Ewald summation method, as previously developed by Kreisel \textit{et al.} \cite{kreiselMicroscopicSpinwaveTheory2009}, and find
\begin{align}
	D_{\kv}^{zz}(z_{ij})&=\frac{\pi\mu^{2}}{a^{2}}\sum_{\gv}
	\left(\frac{8\sqrt{\varepsilon}}{3\sqrt{\pi}}e^{-p^{2}-q^{2}}-|\kv+\gv|f(p,q)\right) \nonumber\\
	&\qquad -\frac{4\mu^{2}}{3}\sqrt{\frac{\varepsilon^{5}}{\pi}}\sum_{\rv}\left(|\rv_{ij}|^{2}-3z_{ij}^{2}\right)\cos\left(k_{x}x_{ij}\right)\cos\left(k_{y}y_{ij}\right)\varphi_{3/2}(|\rv_{ij}|^{2}\varepsilon);\\
	D_{\kv}^{yy}(z_{ij})&=\frac{\pi\mu^{2}}{a^{2}}\sum_{\gv}
	\left(\frac{4\sqrt{\varepsilon}}{3\sqrt{\pi}}e^{-p^{2}-q^{2}}-\frac{(k_{y}+g_{y})^{2}}{|\kv+\gv|}f(p,q)\right)  \nonumber	 \\
	&\qquad -\frac{4\mu^{2}}{3}\sqrt{\frac{\varepsilon^{5}}{\pi}}\sum_{\rv}\left(|\rv_{ij}|^{2}-3y_{ij}^{2}\right)\cos\left(k_{x}x_{ij}\right)\cos\left(k_{y}y_{ij}\right)\varphi_{3/2}(|\rv_{ij}|^{2}\varepsilon); \\
	D_{\kv}^{xy} (z_{ij}) &= -\frac{\pi \mu^2} {a^2} \sum_{\gv}
	\frac{(k_y + g_y) (k_x + g_x)}{|\kv+\gv|}  f(p, q) \nonumber\\
	&\qquad-4 \frac{\varepsilon^{5 / 2} \mu^2}{\sqrt{\pi}} \sum_{\rv}  y_{ij} x_{ij} 
	\sin(k_x  x_{ij})  \sin(k_y  y_{ij})  \varphi_{3/2}(|\rv_{ij}|^{2}\varepsilon),
\end{align}
where 
\begin{equation}
	\varphi_{3/2}(x) = e^{-x} \frac{3 + 2x}{2x^2} + \frac{3\sqrt{\pi} \erfc \left( \sqrt{x} \right)}{4x^{5/2}}
\end{equation}
and $q=z_{ij}\sqrt{\varepsilon}$, $p=|\kv + \bm{g}|/(2\sqrt{\varepsilon})$ and
$f(p,q)=e^{-2pq}\erfc(p-q) + e^{2pq}\erfc(p+q)$.
The sums are either over the real space lattice or the reciprocal lattice, where the reciprocal lattice vectors are $g_x=2\pi m$, $g_y=2\pi n$, $\{m,n\}\in \mathbb{Z}$.
$\varepsilon$ determines the ratio between the reciprocal and real sums. We choose $\varepsilon=a^{-2}$, such that $2pq \approx 1$ and $\exp[\pm 2pq]$ converges quickly.  
Note that $D_{\kv}^{xx}$ from the symmetry $D_{\kv}^{yy}=D_{\kv}^{xx}(k_{x}\rightarrow k_{y},k_{y}\rightarrow k_{x})$. Taking the derivatives w.r.t. $k_x$ and $k_y$ we find
\begin{align}
	\partial_{k_y} D_{\kv}^{zz}(z_{ij})&=\frac{\pi\mu^{2}}{a^{2}}\sum_{\gv}
	\left(\frac{16p\sqrt{\varepsilon}}{3\sqrt{\pi}}e^{-p^{2}-q^{2}} \frac{\partial p}{\partial k_y} + p2\sqrt{\varepsilon} \frac{\partial f}{\partial k_y} + \frac{k_y + g_y}{2\sqrt{\varepsilon}p} f(p,q)\right) \nonumber\\
	&\qquad +\frac{4\mu^{2}}{3}\sqrt{\frac{\varepsilon^{5}}{\pi}}\sum_{\rv} y_{ij}\left(|\rv_{ij}|^{2}-3z_{ij}^{2}\right)\cos\left(k_{x}x_{ij}\right)\sin\left(k_{y}y_{ij}\right)\varphi_{3/2}(|\rv_{ij}|^{2}\varepsilon);\\
	\partial_{k_y} D_{\kv}^{yy}(z_{ij})&=-\frac{\pi\mu^{2}}{a^{2}}\sum_{\gv}
	\left(\frac{8p\sqrt{\varepsilon}}{3\sqrt{\pi}}e^{-p^{2}-q^{2}}\frac{\partial p}{\partial k_y}
	+\frac{\left(k_y + g_y\right)^2}{|\kv+\gv|} \frac{\partial f}{\partial k_y}
	+\frac{2(k_{y}+g_{y})|\kv+\gv|^2 - (k_{y}+g_{y})^3}{|\kv+\gv|^3}f(p,q)\right)  \nonumber	 \\
	&\qquad +\frac{4\mu^{2}}{3}\sqrt{\frac{\varepsilon^{5}}{\pi}}\sum_{\rv}y_{ij}\left(|\rv_{ij}|^{2}-3y_{ij}^{2}\right)\cos\left(k_{x}x_{ij}\right)\sin\left(k_{y}y_{ij}\right)\varphi_{3/2}(|\rv_{ij}|^{2}\varepsilon); \\
	\partial_{k_x} D_{\kv}^{yy}(z_{ij})&=-\frac{\pi\mu^{2}}{a^{2}}\sum_{\gv}
	\left(\frac{8p\sqrt{\varepsilon}}{3\sqrt{\pi}}e^{-p^{2}-q^{2}}\frac{\partial p}{\partial k_x} 
	-\frac{(k_{y}+g_{y})^{2}}{|\kv+\gv|} \frac{\partial f}{\partial k_x} + \frac{(k_y + g_y)^2 (k_x + g_x)}{|\kv+\gv|^3} f(p,q)  \right) \nonumber	 \\
	&\qquad +\frac{4\mu^{2}}{3}\sqrt{\frac{\varepsilon^{5}}{\pi}}\sum_{\rv}z_{ij}\left(|\rv_{ij}|^{2}-3y_{ij}^{2}\right)\cos\left(k_{y}y_{ij}\right)\sin\left(k_{z}z_{ij}\right)\varphi_{3/2}(|\rv_{ij}|^{2}\varepsilon); \\
	\partial_{k_y} D_{\kv}^{xy} (z_{ij}) &= -\frac{\pi \mu^2} {a^2} \sum_{\gv}
	\frac{(k_y + g_y) (k_x + g_x)}{|\kv+\gv|}  \frac{\partial f}{\partial k_y} +  \frac{(k_x+g_x)|\kv+\gv|^2 - (k_y+g_y)^2(k_x+g_x)}{|\kv+\gv|^3} f(p,q) \nonumber\\
	&\qquad-4 \frac{\varepsilon^{5 / 2} \mu^2}{\sqrt{\pi}} \sum_{\rv}  y_{ij}^2 x_{ij} 
	\sin(k_x  x_{ij})\cos(k_y  y_{ij})    \varphi_{3/2}(|\rv_{ij}|^{2}\varepsilon),
\end{align}
where 
\begin{gather}
	\frac{\partial p}{\partial k_\alpha} = \frac{k_\alpha + g_\alpha}{4\varepsilon p};\\
	\frac{\partial f}{\partial k_\alpha} = \left( 2q e^{2pq}\erfc({p+q}) - 2qe^{-2pq}\erfc({p-q}) - \frac{4}{\sqrt{\pi}} e^{-p^2-q^2}\right)\frac{k_\alpha + g_\alpha}{4p\varepsilon}
\end{gather}
and the remaining terms follow from symmetry, by swapping $k_y\leftrightarrow k_x$.

\end{widetext}

\onecolumngrid

\end{document}